\documentclass[%
 aip,
 jmp,%
 amsmath,amssymb,
 reprint,%
]{revtex4-1}

\usepackage{graphicx}
\usepackage{dcolumn}
\usepackage{bm}
\usepackage{color}

\begin{document}


\title[Stochastic trajectories using Brownian Bridges]
{Sampling constrained stochastic trajectories using Brownian bridges}

\author{Patrice Koehl}
\email{koehl@cs.ucdavis.edu}
\affiliation {Department of Computer Sciences, University of California, Davis, CA 95616, USA}
\author{Henri Orland}
\email{henri.orland@cea.fr}
\affiliation{Universit\'{e} Paris-Saclay, Institut de Physique Th\'{e}orique, CEA, CNRS, F-91191 Gif-sur-Yvette, France}

\date{\today}

\begin{abstract}
We present a new method to sample conditioned trajectories of a system evolving under Langevin dynamics, based on Brownian bridges. 
The trajectories are conditioned to end at a certain point (or in a certain region) in space. 
The bridge equations can be recast exactly in the form of a non linear stochastic integro-differential equation. 
This equation can be very well approximated when the trajectories are closely bundled together in space, i.e. at low temperature, or for transition paths. The approximate equation can be solved iteratively, using a fixed point method. 
We discuss how to choose the initial trajectories and show some examples of the performance of this method
on some simple problems. 
The method allows to generate conditioned trajectories with a high accuracy.
\end{abstract}

\keywords{Langevin equation, Stochastic trajectories, Minimum action path}

\maketitle

 \section{Introduction}

With the availability of extremely powerful computers, stochastic simulations have become the main tool to explore a large variety of physical, chemical and biological phenomena such as the spontaneous folding-unfolding of proteins \cite{Shaw:2008}, allosteric transitions \cite{Elber:2011}, the binding of molecules \cite{Shan:2011}, or more generally to compute physical observables of complex systems \cite{Wang:2020}.
In some cases, however, the duration of the phenomenon under study can be extremely long, still beyond present computational capabilities.
Fortunately, for many of those phenomena, most of the duration is spent waiting for the interesting event to occur. These are called {\em rare events}, and their probability of occurrence is exponentially small.
Protein folding is one such case \cite{Hartmann:2014}. Indeed, although the total folding time may be of the order of seconds, the time during which the system effectively jumps from an unfolded state to the folded state can be much shorter, of the order of microseconds. This most interesting part of the trajectory during which the system effectively evolves from an unfolded state to the folded state is called the {\em transition path}. It has been shown that the typical time between folding-unfolding events is given by the Kramers time \cite{Hanggi:1990}, which is exponential in the barrier height, whereas the duration of the folding itself , called the {\em transition path time} is logarithmic in the barrier height \cite{Szabo,Zhang:2007,Laleman:2017}.

\begin{figure}[htbp]
\centering
\includegraphics[width=.6\hsize]{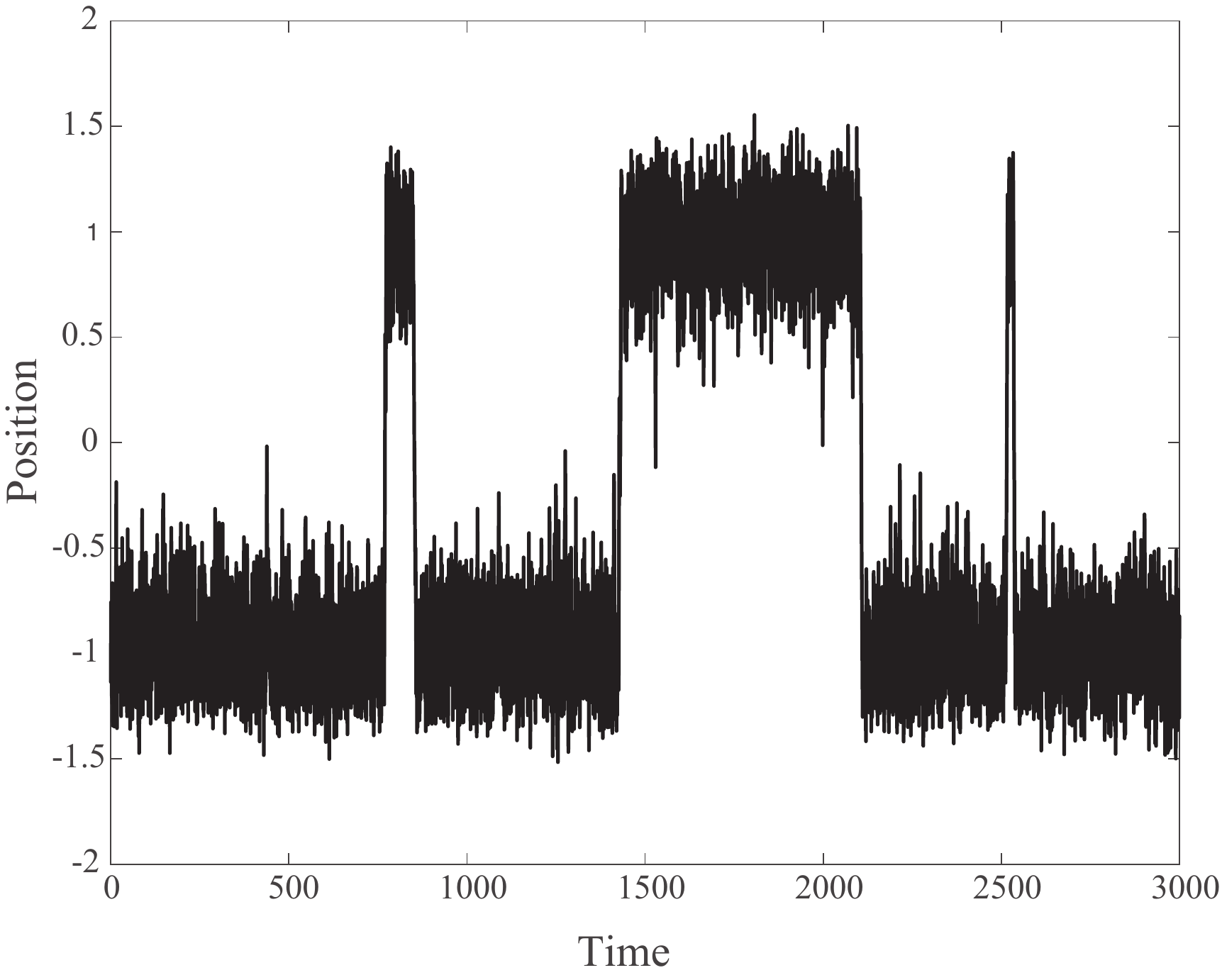}
\caption{Stochastic trajectory (see equation \ref{eq:langevin} of a particle in a double well potential, $U=0.25(x^2-1)^2$. The trajectory is run at a temperature $T=0.01$, with a step size in time $dt=0.1$, over 20000 steps. Note that the transition paths (almost vertical jumps) are very short compared to the sampling of the bottom of the two wells. }
\label{quartic}
\end{figure}

The archetype of such behavior is illustrated in the tunnelling of a classical particle in a quartic potential (see figure \ref{quartic}), under thermal noise 

Single molecule experiments have triggered a renewed interest in the study of these transition trajectories \cite{Chung:2018, Neupane:2012}.
It is thus desirable to be able to follow the dynamics of the system during this transition path time, and monitor the large conformational changes undergone by the system. These studies would also allow for a microscopic determination of the {\em transition state} and of the barrier height of the transition. Such knowledge is being used in drug design, in trying to modify the barrier height or to block the transition by binding to the transition state \cite{Spagnolli:2021}.

In this article we deal with the problem of sampling the stochastic trajectories of a system, which starts in a certain known configuration at the initial time, and transitions to a known final state (or family of final states) in a given time $t_f$. 
The goal is to sample the family of such transition trajectories.

There has been numerous works on this problem (for a review see \cite{Elber:2020}), originating with the {\em transition state theory} (TST) \cite{Eyring:1935, Wigner:1938}.
In the TST, the transition state is identified with a saddle-point of the energy surface, and the most probable transition path is set as the minimum energy path (MEP) along that surface. Due to the limitations of the TST to smooth energy surfaces, E and Vanden-Eijnden \cite{E:2002} have developed the {\em string method} which is exact at zero temperature and provides a framework for finding the most probable transition path between two conformations of a molecule. This framework has served as the touchstone for many path finding algorithms, such as the string method, the nudged elastic band method, and others (see Ref. \cite{Elber:2020}).
Other algorithms were developed to minimize or even sample the Onsager-Machlup action, in order to generate the most probable path or its neighbourhood \cite{Olender:1996, Eastman:2001, Franklin:2007, Faccioli:2006}.

This paper sits as a continuation of our recent work on the bridge equation, a  modified Langevin equation that is conditioned to join the two predefined end states of the system under study \cite{Orland:2011, Majumdar:2015, Delarue:2017}.  
We first revisit this concept , as described in the following section, and show that it can be recast exactly in the form of a non linear
stochastic integro-differential equation. We show that this equation can be very well approximated when the trajectories are closely bundled together in space, i.e. at low temperature. We then derive a fixed point method to solve this equation efficiently.
Finally, we illustrate the method on simple cases, a quartic potential and the classical problem of finding minimum energy trajectories along the Mueller potential \cite{Mueller:1979, Mueller:1980}.

\section{Theory}

\subsection{An integro-differential form of the bridge equation}

Consider a system of particles, with $N$ degrees of freedom, represented
by a position vector ${\bf r} \in \mathbb{R}^3$. The particles of the system interact through
a conservative force derived from a potential $U$. The system is
evolved using overdamped Langevin dynamics
\begin{equation}
\dot{{\bf r}}=-\frac{1}{\gamma}{\bf \nabla}U+\bm{\eta}(t)
\label{eq:langevin}
\end{equation}
where ${\bf F}=-\nabla$U is the force acting on the system, $\bm{\eta}$
is the Gaussian random force, and $\gamma$ is the friction coefficient.
The friction coefficient is related to the diffusion constant $D$
and the temperature $T$ through the Einstein relation
\begin{equation}
\gamma=\frac{k_{B}T}{D}=\frac{1}{D\beta}\label{eq:einstein}
\end{equation}
where $\beta=1/k_{B}T$. The friction is usually taken to be independent
of $T$ , so that the diffusion coefficient $D$ is proportional to
the temperature $T$.

The moments of the Gaussian white noise are given by
\begin{eqnarray}
\langle\eta_{a}(t) \rangle & = & 0\nonumber \\
\langle\eta_{a}(t)\eta_{a'}(t')\rangle & = & 2D\delta_{aa'}\delta(t-t')\label{eq:noise}
\end{eqnarray}
where the indices $a$ and $a'$ denote components of the vector $\bm{\eta}(t)$.
As the diffusion constant $D$ is proportional to $T$, the random force $\bm{\eta}(t)$ is
of order $\sqrt{T}$.

The probability distribution function $P({\bf r},t|{\bf r}_{i},0)=P({\bf r},t)$
for the system to be at position ${\bf r}$ at time $t$ given that
it was at position $\bm{r}_{i}$ at time 0, satisfies the Fokker-Planck
(FP) equation
\begin{equation}
\frac{\partial P}{\partial t}=D\nabla\left(\nabla P+\beta\nabla U({\bf r})P\right)\label{eq:FP}
\end{equation}
Among all the paths generated by the Langevin equation (\ref{eq:langevin}),
we are only interested in those which are conditioned to end at a given point
${\bf r}_{f}$ at time $t_{f}$.
As a side note, one could treat in the same manner paths conditioned to end in a certain region of space at $t_f$. 
Although these paths are in general
of zero measure in the ensemble of paths originating from $({\bf r}_{i},0)$
, there is an infinite number of them. We are interested to
generate only those paths satisfying this constraint. For this purpose,
we use the method of Brownian bridges introduced through the Doob transform \cite{Doob:1957}. We denote by $\mathcal{P}({\bf r},t)$
the probabilty that the conditioned system is at point ${\bf r}$
at time $t$. We have
\begin{equation}
\mathcal{P}({\bf r},t)=\frac{P({\bf r}_{f},t_{f}|{\bf r},t)P({\bf r},t|{\bf r}_{i},0)}{P({\bf r}_{f},t_{f}|{\bf r}_{i},0)}\label{eq:proba}
\end{equation}
The probability $P({\bf r},t|{\bf r}_{i},0)$ satisfies eq.(\ref{eq:FP})
whereas the function $Q_{1}({\bf r},t)=P({\bf r}_{f},t_{f}|{\bf r},t)$ above 
satisfies the reverse or adjoint Fokker-Planck (FP) equation \cite{VanKampen:1992}.

\begin{equation}
\frac{\partial Q_{1}}{\partial t}=-D\nabla^{2}Q_{1}+D\beta\nabla U({\bf r})\nabla Q_{1}\label{eq:FPadjoint}
\end{equation}
Using eq.(\ref{eq:FP}) and (\ref{eq:FPadjoint}), one can easily
see that $\mathcal{P}({\bf r},t)$ satisfies the modified FP equation

\begin{equation}
\frac{\partial\mathcal{P}}{\partial t}=D\nabla\left(\nabla\mathcal{P}+\nabla\left(\beta U({\bf r})-2\ln Q_{1}\right)P\right)\label{eq:modifFP}
\end{equation}
from which we see that the position ${\bf r}(t)$ of the conditioned
system satisfies a modified Langevin equation given by
\begin{equation}
\dot{{\bf r}}=-\frac{1}{\gamma}{\bf \nabla}U+2D\nabla\ln Q_{1}+\bm{\eta}(t)\label{eq:bridge}
\end{equation}
This equation is called a bridge equation \cite{Doob:1957}. The additional force
term $2D\nabla\ln Q_{1}$ in the Langevin equation conditions the paths and guarantees that
they will end at $({\bf r}_{f},t_{f})$. We can use a path integral
representation for $Q_{1}$ \cite{Feynman:1965},
\begin{eqnarray}
Q_{1}({\bf r},t) & = & P({\bf r}_{f},t_{f}|{\bf r},t)\nonumber \\
 & = & \int_{{\bf r}(t)={\bf r}}^{{\bf r}(t_{f})={\bf r}_{f}}\mathcal{D}{\bf r}(\tau)e^{-\frac{1}{4D}\int_{t}^{t_{f}}d\tau\left(\dot{{\bf r}}+\frac{1}{\gamma}{\bf \nabla}U\right)^{2}}\label{eq:ito}\\
 & = & e^{-\beta\left(U({\bf r}_{f})-U({\bf r})\right)} \times \nonumber \\
 &&\int_{{\bf r}(t)={\bf r}}^{{\bf r}(t_{f})={\bf r}_{f}}\mathcal{D}{\bf r}(\tau)e^{-\int_{t}^{t_{f}}d\tau\left(\frac{\dot{{\bf r}}^{2}}{4D}+\frac{1}{D\gamma^{2}}V({\bf r(\tau))}\right)} \nonumber \\\label{eq:strato} \\
 &=&e^{-\beta\left(U({\bf r}_{f})-U({\bf r})\right)} \langle {\bf r} \left| e^{-(t_f - t) H} \right | {\bf r} \rangle.
  \label{QM}
\end{eqnarray}
In eq.(\ref{QM}), we have used standard quantum mechanical notation \cite{Feynman:1965} for the matrix element of the evolution operator $e^{-Ht}$ ,  where the Hamiltonian $H$ is given by
\begin{equation}
H=-D\nabla^2+ D\beta^2 V({\bf r}) 
\label{eq:hamiltonien}
\end{equation}
and the effective potential $V$ is given by
\begin{equation}
V({\bf r})=\frac{1}{4}\left(\nabla U\right)^{2}-\frac{k_{B}T}{2}\nabla^{2}U\label{eq:effective}
\end{equation}
The driving term $Q_{1}({\bf r},t)$ is a sum over all paths joining $({\bf r},t)$
to $({\bf r}_{f},t_{f}),$ properly weighted by the so-called Onsager-Machlup
action \cite{Onsager:1953},  $\frac{1}{4D}\int_{t}^{t_{f}}d\tau\left(\dot{{\bf r}}+\frac{1}{\gamma}{\bf \nabla}U\right)^{2}$.

The above equations are obtained by transforming the Ito form of the
path integral (\ref{eq:ito}) into the Stratonovich form (\ref{eq:strato}),
when expanding the square, and using the identity of stochastic calculus
\cite{Stratonovich:1971, Elber:2020}
\begin{equation}
\int_{t}^{t_{f}}d\tau\dot{{\bf r}}{\bf \nabla}U({\bf r}(\tau))=U({\bf r}_{f})-U({\bf r})-D\int_{t}^{t_{f}}d\tau\nabla^{2}U({\bf r}(\tau))\label{eq:integral}
\end{equation}
Defining
\begin{equation}
Q({\bf r},t)=\int_{{\bf r}(t)={\bf r}}^{{\bf r}(t_{f})={\bf r}_{f}}\mathcal{D}{\bf r}(\tau)e^{-\int_t^{t_f} d\tau\left(\frac{\dot{{\bf r}}^{2}}{4D}+\frac{1}{D\gamma^{2}}V({\bf r(\tau))}\right)}\label{eq:Q}
\end{equation}
the bridge equation (\ref{eq:bridge}) becomes
\begin{equation}
\dot{{\bf r}}=2D\nabla\ln Q+\bm{\eta}(t)\label{eq:redbridge}
\end{equation}

Using the path integral representation (\ref{eq:Q})  and performing several integrations by part, we show in Appendix A that this equation can be exactly recast in the following form
\begin{equation}
\dot{{\bf {r}}}=\frac{{\bf r}_{f}-{\bf r}(t)}{t_{f}-t}-\frac{2}{\gamma^{2}}\int_{t}^{t_{f}}d\tau\left(\frac{t_{f}-\tau}{t_{f}-t}\right)\langle\nabla V({\bf r}(\tau))\rangle_Q+\bm{\eta}(t)
\label{eq:star}
\end{equation}
where the bracket $\langle\cdots\rangle _Q$ denotes the average over
all paths joining $({\bf r},t)$ to $({\bf r}_{f},t_{f}),$ weighted
by the action of eq. (\ref{eq:Q})
\begin{widetext}
\begin{eqnarray}
\label{eq:average}
\langle\nabla V({\bf r}(\tau))\rangle_Q&=&\frac{1}{Q({\bf r},t)}\int_{{\bf r}(t)={\bf r}}^{{\bf r}(t_{f})={\bf r}_{f}}\mathcal{D}{\bf r}(\tau)\nabla V({\bf r}(\tau))\,e^{-\int_{t}^{t_{f}}d\tau\left(\frac{\dot{{\bf r}}^{2}}{4D}+\frac{1}{D\gamma^{2}}V({\bf r(\tau))}\right)} 
= \frac{\langle {\bf r}_f \left| e^{-(t_f - \tau) H} \nabla V({\bf r}) e^{-(\tau-t)H} \right | {\bf r} \rangle}{\langle {\bf r}_f \left| e^{-(t_f - t) H} \right | {\bf r} \rangle}
\end{eqnarray}
\end{widetext}
and the Gaussian noise is defined by eq.(\ref{eq:noise}).
Note
that the first term in the r.h.s of equation (\ref{eq:star}) guarantees that
the constraint ${\bf r}(t_{f})={\bf r}_{f}$ is satisfied. It is the only term which is singular at time $t_f$, since the integral term does not have any singularity at any time.
In fact, in the case of a free Brownian particle, the effective potential $V$ vanishes, and we recover the standard equation for free Brownian bridges
\begin{equation}
\dot{{\bf {r}}}=\frac{{\bf r}_{f}-{\bf r}(t)}{t_{f}-t}+\bm{\eta}(t)\label{eq:free}
\end{equation}

Equation (\ref{eq:star}) is the fundamental equation of
this article and will be used to generate constrained paths.
This equation 
is a non linear stochastic equation. It is Markovian, in the sense that the right hand side of (\ref{eq:star}) depends only on ${\bf r}(t)$. However, the presence of the average over all future paths makes it difficult to use. 

\subsection{Zero temperature and low temperature expansions of the bridge equation}

At zero temperature, the noise term vanishes and the average in (\ref{eq:star})
reduces to a single trajectory ${\bf r}_{0}(t)$. The equation becomes

\begin{equation}
\dot{{\bf r}}_{0}=\frac{{\bf r}_{f}-{\bf r}_{0}(t)}{t_{f}-t}-\frac{2}{\gamma^{2}}\int_{t}^{t_{f}}d\tau\left(\frac{t_{f}-\tau}{t_{f}-t}\right)\nabla V_{0}({\bf r}_{0}(\tau))\label{eq:zero}
\end{equation}
where $V_{0}({\bf r})=\frac{1}{4}\left(\nabla U\right)^{2}$ is the
zero temperature effective potential (see eq. \ref{eq:effective}). Let us
show that this equation is equivalent to the usual zero temperature
instanton equation \cite{Zinn:2002}. Indeed, taking a time derivative of the
above equation we get easily

\begin{equation}
\ddot{{\bf r}}_{0}=\frac{2}{\gamma^{2}}\nabla V_{0}({\bf r}_{0})=\frac{1}{\gamma^{2}}\nabla U({\bf r}_{0}).\nabla^{2}U({\bf r}_{0})\label{eq:instanton}
\end{equation}
which, supplemented by the boundary conditions ${\bf r}_{0}(0)={\bf r}_{i}$
and ${\bf r}_{0}(t_{f})={\bf r}_{f}$, is the standard instanton equation. As will be discussed later, the non-linear equation (\ref{eq:zero})
can be solved iteratively, starting from an initial trajectory. This method provides an efficient algorithm to compute the zero temperature trajectory connecting the two points. 
The two equivalent equations (\ref{eq:zero})
and (\ref{eq:instanton}) can have multiple solutions which can be obtained by modifying the initial
guessed trajectory. 

One can perform a low temperature expansion of eq.(\ref{eq:star}) around the zero temperature trajectory ${\bf r_0}(t)$. The lowest order in $T$ is of order $\sqrt{T}$ and the equation is simply obtained by adding the noise term
\begin{equation}
\dot{{\bf r}}=\frac{{\bf r}_{f}-{\bf r}(t)}{t_{f}-t}-\frac{2}{\gamma^{2}}\int_{t}^{t_{f}}d\tau\left(\frac{t_{f}-\tau}{t_{f}-t}\right)\nabla V_{0}({\bf r}(\tau))
+ {\bm \eta}(t) 
\label{eq:low}
\end{equation}
Similarly to eq.(\ref{eq:zero}), this equation is to be solved self-consistently. We will propose below a fixed point solution to this problem.

\subsection{Perturbative Expansion}
%
A perturbative expansion in powers of the effective potential $V$ can be easily obtained from eq.(\ref{eq:star}) by expanding the average (\ref{eq:average}). To lowest order in $V$, 
eq.(\ref{eq:star}) becomes
\begin{eqnarray}
\label{cumulant}
\dot{{\bf {r}}}&=&\frac{{\bf r}_{f}-{\bf r}(t)}{t_{f}-t}-\frac{2}{\gamma^{2}} \nonumber \\
&&\int_{t}^{t_{f}}d\tau\left(\frac{t_{f}-\tau}{t_{f}-t}\right)
\int_{-\infty}^{+\infty} \frac{d {\bf z}}{ (2\pi)^{N/2}} e^{-\frac{{\bf z}^2}{2}} \nabla V({\bf R}(\tau))
+\bm{\eta}(t) \nonumber \\
\end{eqnarray}
where
\begin{equation}
\label{R}
{\bf R}(\tau) = \frac{{\bf r}_f (\tau-t)+{\bf r}(t) (t_f-\tau)}{t_f-t} + \sqrt{\frac{2 D (t_f-\tau)(\tau-t)}{t_f-t}} {\bf}{\bf z}
\end{equation}
and $\bf z$ is an $N$-dimensional Gaussian vector with zero mean and variance 1. The interpretation of this approximation with respect to eq.(\ref{eq:star}) is quite simple: the ensemble of trajectories from $({\bf r(t)}, t)$ to $({\bf r}_f, t_f)$  which enter the expectation value over $Q$, is approximated by a straight line to which a Gaussian random noise with variance $\sigma(\tau)= \sqrt{\frac{2 D (t_f-\tau)(\tau-t)}{t_f-t}}$ is added and the average over the ensemble $Q$ is approximated by a Gaussian integral on $\bf z$.
This equation is identical to the one obtained by the cumulant expansion discussed in Ref. \cite{Delarue:2017} where it has been shown to describe accurately the dynamics of the transitions.
\subsection{An efficient approximation for weakly dispersed trajectories}
Consider eq.(\ref{eq:star}) and take the average over all future noise history as prescribed by $\langle \ldots \rangle_Q$. 
If the trajectories are weakly dispersed around an average trajectory (which is the case at low temperature or for transition paths), we can use the approximation
\begin{eqnarray}
\label{narrow}
\nabla \langle V({\bf r}(\tau))\rangle_Q  &\sim&  \nabla V( \langle{\bf r}(\tau) \rangle_Q) \nonumber \\
\end{eqnarray}
Eq.(\ref{eq:star}) becomes
\begin{equation}
\dot{{\bf {r}}}=\frac{{\bf r}_{f}-{\bf r}(t)}{t_{f}-t}-\frac{2}{\gamma^{2}}\int_{t}^{t_{f}}d\tau\left(\frac{t_{f}-\tau}{t_{f}-t}\right)\nabla V({\bf r}(\tau))+\bm{\eta}(t)
\label{eq:newstar1}
\end{equation}
in which we have replaced $\langle {\bf r}(\tau) \rangle_Q$ by ${\bf r}(\tau)$ in the integral term. The argument for this replacement is that if all trajectories ${\bf r}(t)$ are bunched together, they are all close to the average trajectory ${\langle {\bf r}(t)\rangle_Q}$ and one can use the further approximation ${\bf r}(t) \approx {\langle {\bf r}}(t)\rangle_Q$. Note that the above eq.(\ref{eq:newstar1}) is very similar to eq.(\ref{eq:low}), except for the replacement of $V_0$ by $V$ in (\ref{eq:low}). 
In particular, it is exact at zero temperature.

All the relevant approximate equations discussed above are non-linear integro-differential equations, in which the evolution of ${\bf r}(t)$ depends on the trajectory at future times $\tau >t$. 
 We now discuss how to solve these equations.

\subsection{A fixed point method for solving the approximate integro differential bridge equations}

The simplest method to solve the non-linear integro-differential stochastic
equations (\ref{eq:zero}), (\ref{eq:low}), and (\ref{eq:newstar1})
is to use an iterative fixed point method.

There are several ways to implement the iterative method to solve
these equations, depending on the way one splits them into the form of a recursion.
In addition, the convergence of the method depends crucially on the
choice of the initial guess. On the examples we studied, we found
that a simple method to solve the equation is to use a Euler-Maruyama \cite{Kloeden:1992}
discretization scheme for the equation, dividing the time $t_{f}$
in $I$ intervals of size $dt$, so that $t_{f}=Idt$. The integral
can be calculated with the same $dt$ or with a larger one, to speed
up the computation. 
Using the example of eq.(\ref{eq:newstar1}) and 
denoting by ${\bf r} ^{(n)}(k)$ the $n$-th iteration
of the trajectory at time $t=kdt$, we write the iteration as 
\begin{widetext}
\begin{equation}
{\bf r}^{(n+1)}(k+1)={\bf r}^{(n)}(k)dt+\frac{{\bf r}_{f}-{\bf r}^{(n)}(k)}{t_{f}-kdt}dt-\frac{2 dt}{\gamma^{2}}\sum_{k'=k}^{I-1}\left(\frac{t_{f}-k'dt}{t_{f}-kdt}\right)dt\nabla V({\bf r}^{(n)}(k'))+\sqrt{2Ddt}\bm{\xi}(k)
\label{eq:discret}
\end{equation}
\end{widetext}
where $\bm{\xi}(k)$ is a normalized Gaussian variable:
\begin{eqnarray}
\langle\xi_{a}(k)\rangle & = & 0; \quad \quad \langle\xi_{a}^{2}(k)\rangle = 1.
\end{eqnarray}
In those equations $\xi_{a}(k)$ is the $a-$th component of $\bm{\xi}(k)$.
In equation (\ref{eq:discret}), the initial condition is ${\bf r}^{(n)}(0)={\bf r}_{i}$
and the noise $\bm{\xi}(k)$ is the same for all the iterations. This
equation is iterated in $n$ until convergence.
As stated above, the convergence of the process very much depends
on the initial guessed trajectory $\{{\bf r}^{(0)}(k)\}$.
We now discuss two possible choices of initial trajectories:
\begin{itemize}
\item \textbf{\emph{Brownian Bridges}}.
This is the simplest method, and it works very well in many  tested cases. The idea is to initialize the iterations with the Brownian Bridge trajectories, generated by eq.(\ref{eq:free}) with the correct boundary conditions. This is extremely fast to implement, and very efficient. It may however violate steric constraints since it does not include the interactions. To circumvent this difficulty, we may use the cumulant expansion.

\item \textbf{\emph{Cumulant expansion trajectories}}.
The cumulant expansion equation (\ref{cumulant}) obtained as a perturbative expansion of equation  (\ref{eq:star}) is a good approximation \cite{Delarue:2017} and thus it is natural to use its trajectories as initial guesses for the fixed point method. In addition, it does not violate steric constraints since it fully takes into account the potential.
\end{itemize}

\section{Results}

We illustrate the use of equations (\ref{eq:zero}) and (\ref{eq:newstar1}) on two simple examples, namely the double well quartic potential in one dimension, and the Mueller potential in two dimensions. In both cases, initial trajectories are taken as simple Brownian bridges.

\subsubsection*{ Quartic potential}

We study first the one-dimensional double-well potential
\begin{equation}
U(x)=\frac{1}{4}(x^2-1)^2
\end{equation}
It has a barrier of energy $\Delta E=0.25$ between the two minima at $\pm1$.
The effective potential is easily computed from the derivatives of $U$: $V(x)=U'^2(x) - 2k_BT U''(x) = (x^3-x)^2 - 2k_BT(3x^2-1)$.
A key advantage of using this potential is that it is simple enough that we can solve directly the Bridge equation \ref{eq:bridge} (see \cite{Majumdar:2015} for details) and compare the corresponding ``exact" solution with the solution obtained with the approximation proposed in this paper.

A trajectory between minima of a given potential is defined by 3 parameters: its total duration, $t_f$, the time step used to discretize $t_f$, $dt$, and the ambient temperature, $T$. As shown in Ref. \cite{Laleman:2017}, the average transition path time for the quartic potential $\tau_{TP}$ varies as $\log \beta \Delta E$ and depends weakly on the temperature. It is typically $\tau_{TP} \approx 2$. We have therefore studied transition paths between the minima at $\pm 1$ over a total time $t_f = 3$. We chose a time step $dt=10^{-3}$, i.e. a discretization of the total time with $3000$ steps.

\begin{figure}[htbp]
\centering
\includegraphics[width=.5\hsize]{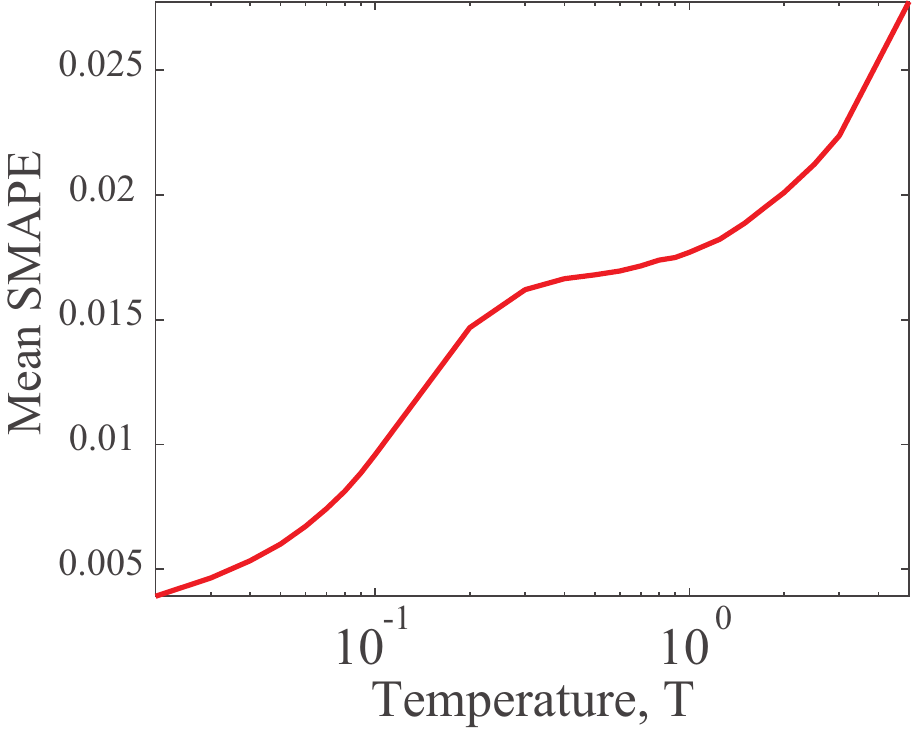}
\caption{Mean symmetric mean absolute percentage error scores between exact and approximate trajectories for the quartic potential as a function of the temperature (see text for details)}
\label{figB}
\end{figure}

To illustrate the impact of temperature on the approximation within eq.(\ref{eq:newstar1}), we ran the following set of experiments. We compared exact and approximate
trajectories for varying temperatures from 0.02 to 3. For each temperature, we compared one thousand pairs of exact ${\bf r_e}$ and approximate ${\bf r_a }$trajectories. The trajectories within
each pair are computed with the same noise sequences. The approximate trajectory based on eq.(\ref{eq:newstar1}) is initialized with a Brownian bridge, followed by iterations of  eq.(\ref{eq:discret}) until convergence, i.e. when  the trajectories at iterations $n$ and $n+1$ differ by less than a tolerance $tol$ set to $10^{-6}$. (usually within 5000 iterations).
The exact trajectory is computed by first diagonalizing the Hamiltonian given by eq. (\ref{eq:hamiltonien}) and then solving the bridge equation  \ref{eq:bridge}) (see \cite{Majumdar:2015}. The approximate and exact trajectories  are  then compared using the symmetric mean absolute percentage error, SMAPE:
\begin{eqnarray}
\text{SMAPE}(\mathbf{r_e}, \mathbf{r_a}) = \frac{200}{M} \sum_{i=1}^M \frac{| r_e(i)-r_a(i) |}{| r_e(i) |+ | r_a(i) |}
\label{eq:mape}
\end{eqnarray}
where $M$ is the number of points in both trajectories. In fig.\ref{figB}, we report the average SMAPE values over the one thousand pairs of trajectories as a function of temperature.
We observe three main regimes.
The agreement between exact and approximate trajectories is excellent at low temperature (up to 0.1) with average SMAPE scores of 0.01 \% or less, really good at medium temperature (between 0.1 and 1), with SMAPE scores between 0.01\% and 0.02\%, and starts to weaken at high temperatures ($> 1$) (although the  average SMAPE scores are still lower than 0.1 \%, i.e indicating that the trajectories remain mostly similar). 

\begin{figure*}[htbp]
\centering
\includegraphics[width=.99\hsize]{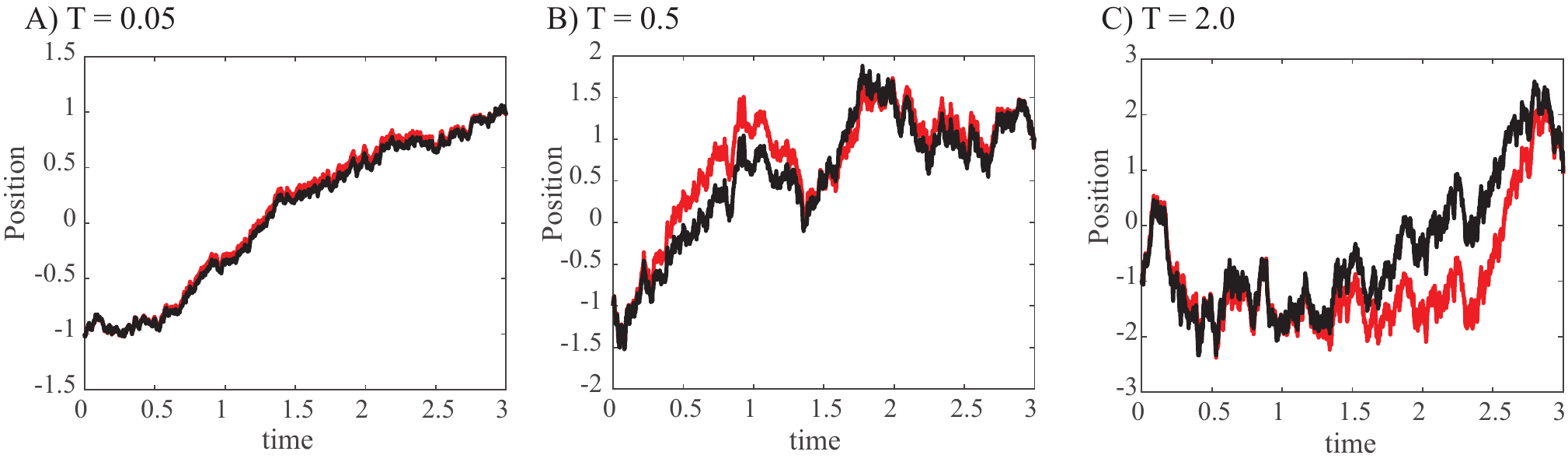}
\caption{Comparisons of the exact trajectories (in black) and the approximate ones computed based on Equation (\ref{eq:newstar1}) (in red)  at  the temperaturesT=0.05 (A), T=0.5 (B) and T=2 (C). The corresponding distances between the exact and approximate trajectories (computed as symmetric mean absolute percentage error, SMAPE (see \ref{eq:mape}) are 0.008\%, 0.017\%, and 0.025 \%, respectively).}
\label{figA}
\end{figure*}

To illustrate those differences, we plot in fig.\ref{figA} three examples of trajectories generated by eq.(\ref{eq:newstar1}) (in red) and by solving exactly the bridge equations (in black) using the same noise, at $T=0.05$ (low temperature), $T=0.5$ (intermediate temperature) and $T=2.$ (high temperature). 

\subsection{The Mueller potential}

\begin{figure*}[htbp]
\centering
\includegraphics[width=.99\hsize]{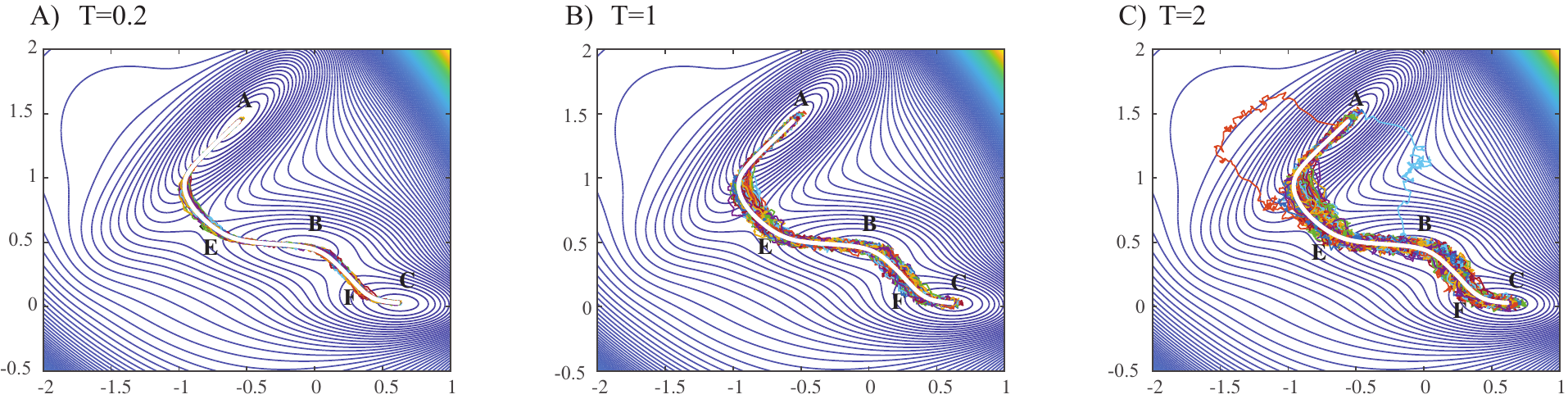}
\caption{Trajectories between the two minima A and C at T=0.2 (A), T=1 (B) and T= 2 (C)}
\label{fig:mueller}
\end{figure*}

The Mueller potential \cite{Mueller:1979, Mueller:1980} is a standard benchmark potential to check the validity of methods for generating transition paths. It is a two-dimensional potential given by
\begin{eqnarray}
&&U(x,y) = \nonumber \\
&& \sum_{i=1}^4 A_i \exp (a_i(x-x_i^0)^2+b_i (x-x_i^0)(y-y_i^0)+c_i(y-y_i^0)^2) \nonumber \\
\end{eqnarray}
with
\begin{eqnarray}
A &=& (-200, -100, -170, 15) \nonumber \\ 
a &=& (-1, -1, -6.5, 0.7) \quad \quad b = (0, 0, 11, 0.6) \nonumber \\
 c &=&(-10, -10, -6,5, 0.7) \nonumber \\
x^0 &=& (1,0,-0,5,-1) \quad \quad y^0  =(0, 0.5, 1.5,1)
\end{eqnarray}
This potential has 3 local minima denoted by A (-0.558, 1.442), B (-0.05, 0.467), and C (0.623, 0.028) separated by two saddle-points at F (-0.793, 0.656) and G(0.198, 0.291) (see for example fig.\ref{fig:mueller}A).


The effective potential $V(x, y)$ can be calculated analytically, as well as its gradient. Equation (\ref{eq:newstar1}) can easily be solved numerically by the fixed point method using eq. (\ref{eq:discret}) . The simulation time $t_f = N_{\rm steps} \ dt$ is chosen so that we observe a small waiting time around the initial as well as the final point. In fig.\ref{fig:mueller}, we display a sample of 100 trajectories starting at A and ending at C,
obtained with $N_{\rm steps}=300$, with a timestep $dt=10^{-4}$ at three temperatures $T =0.2$ (a), $T=1$ (b) and $T=2$ (c). The number of iterations for convergence is around 2000. The zero temperature trajectory is displayed in thick black line in each figure.  

As one can see, at low temperature, the trajectories remain in the vicinity of the zero temperature one, whereas they depart more and more from it as temperature increases.

\section{Conclusion}

In this paper we addressed the problem of generating paths for a system that start at a given  initial configuration and that are conditioned
to end at a given final configuration.
Our approach follows the ideas of Langevin overdamped dynamics, as expressed with  the bridge equation  \cite{Orland:2011, Majumdar:2015, Delarue:2017}.  
We first revisited this concept of bridge between the initial and final configurations and have shown that it can be recast exactly in the form of a non linear
stochastic integro-differential equation. 
We have shown that this equation can be very well approximated when the trajectories are closely bundled together in space, i.e. at low temperature. 
We described one such approximations and we derived a fixed point method to solve the corresponding equations efficiently.
Finally, we illustrated the method on simple test cases, a quartic potential and the classical problem of finding minimum energy trajectories along the Mueller potential.

Our main result is a recast of the bridge equation into a non linear stochastic integro-differential equation, eq. (\ref{eq:star}).
This exact equation is unfortunately difficult to solve, as it expresses the velocity along the trajectory at a time $t$ as an integral of a quantity that is averaged
over the evolutions of trajectories beyond time $t$. However, we have established approximations that proved effective to derive the path at zero temperature,
as well as an ensemble of paths at low temperatures.
Those approximations lead to equations that can be solved iteratively using a fixed point method, see eq. (\ref{eq:discret}).
Solving stochastic differential equations using a fixed point method is not always easy, however, and remains an active research area in numerical analysis.
In this paper, we applied a standard Euler-Maruyama scheme \cite{Kloeden:1992} and showed that it was successful on simple examples, namely a 1D potential and a 2D potential. 
We recognize that we will most likely need more sophisticated solvers for more complicated systems. We are currently working on developing such solvers.

\appendix

\section* {Appendix A}

In this appendix, we prove the central equation of this article, namely
eq.(\ref{eq:star}). For that matter, we need to compute the gradient
of the logarithm of $Q$.

We have

\begin{equation}
Q({\bf r},t)=\int_{{\bf r}(t)={\bf r}}^{{\bf r}(t_{f})={\bf r}_{f}}\mathcal{D}{\bf r}(\tau)e^{-\int_{t}^{t_{f}}d\tau\left(\frac{\dot{{\bf r}}^{2}}{4D}+\frac{1}{D\gamma^{2}}V({\bf r(\tau))}\right)}\label{eq:Qapp}
\end{equation}
which we discretize by splitting the time interval $t_{f}-t$ in $N-k$
intervals of size $dt$. 
\begin{equation*}
t_{k}=t<t_{k+1}=t+dt<t_{k+2}<\ldots<t_{N-1}<t_{N}=t_{f}
\end{equation*}
with
\begin{equation*}
t_i = t+ (i-k)dt
\end{equation*}
for all $i \in [k, N]$ and
\begin{equation*}
t_{f}-t=(N-k)dt.
\end{equation*}
We write
\begin{eqnarray*}
Q({\bf r},t)=&&\int_{{\bf r}_{k}={\bf r}}^{{\bf r}_{N}={\bf r}_{f}}d{\bf r}_{k+1}\ldots d{\bf r}_{N-1} \\
&&e^{\left[-\sum_{l=k}^{N-1}\left(\frac{\left({\bf r}_{l+1}-{\bf r}_{l}\right)^{2}}{4Ddt}+\frac{dt}{D\gamma^{2}}V({\bf r}_{l})\right)\right]}
\end{eqnarray*}
We have

\begin{widetext}
\begin{eqnarray}
 && \nabla Q({\bf r},t) \nonumber \\
 &=& \int_{{\bf r}_{k}={\bf r}}^{{\bf r}_{N}={\bf r}_{f}}d{\bf r}_{k+1}\ldots d{\bf r}_{N-1}e^{\left[-\sum_{l=k+1}^{N-1}\left(\frac{\left({\bf r}_{l+1}-{\bf r}_{l}\right)^{2}}{4Ddt}+\frac{dt}{D\gamma^{2}}V({\bf r}_{l})\right)\right] }
 \nabla e^{\left[-\left(\frac{\left({\bf r}_{k+1}-{\bf r}\right)^{2}}{4Ddt}+\frac{dt}{D\gamma^{2}}V({\bf r})\right)\right]}\nonumber \\
 & = & \int_{{\bf r}_{k}={\bf r}}^{{\bf r}_{N}={\bf r}_{f}}d{\bf r}_{k+1}\ldots d{\bf r}_{N-1}e^{\left[-\sum_{l=k+1}^{N-1}\left(\frac{\left({\bf r}_{l+1}-{\bf r}_{l}\right)^{2}}{4Ddt}+\frac{dt}{D\gamma^{2}}V({\bf r}_{l})\right)\right]}
  \left[\frac{{\bf r}_{k+1}-{\bf r}}{2Ddt}-\frac{dt}{D\gamma^{2}}\nabla V({\bf r})\right]\times e^{\left[-\left(\frac{\left({\bf r}_{k+1}-{\bf r}\right)^{2}}{4Ddt}+\frac{dt}{D\gamma^{2}}V({\bf r})\right)\right]}\nonumber \\
 & = & \int_{{\bf r}_{k}={\bf r}}^{{\bf r}_{N}={\bf r}_{f}}d{\bf r}_{k+1}\ldots d{\bf r}_{N-1} e^{\left[-\sum_{l=k+1}^{N-1}\left(\frac{\left({\bf r}_{l+1}-{\bf r}_{l}\right)^{2}}{4Ddt}+\frac{dt}{D\gamma^{2}}V({\bf r}_{l})\right)\right]}
  \left[-\nabla_{{\bf r}_{k+1}}-\frac{dt}{D\gamma^{2}}\nabla V({\bf r})\right]\times e^{\left[-\left(\frac{\left({\bf r}_{k+1}-{\bf r}\right)^{2}}{4Ddt}+\frac{dt}{D\gamma^{2}}V({\bf r})\right)\right]}\nonumber \\
  \label{eq:1}
\end{eqnarray}
\end{widetext}
We may then integrate by part the term $\nabla_{{\bf r}_{k+1}}$ and
obtain

\begin{widetext}
\begin{eqnarray}
\nabla Q({\bf r},t) & = & 
\int_{{\bf r}_{k}={\bf r}}^{{\bf r}_{N}={\bf r}_{f}}d{\bf r}_{k+1}\ldots d{\bf r}_{N-1} e^{\left[-\sum_{l=k+2}^{N-1}\left(\frac{\left({\bf r}_{l+1}-{\bf r}_{l}\right)^{2}}{4Ddt}+\frac{dt}{D\gamma^{2}}V({\bf r}_{l})\right)\right]} \nonumber \\
&\times &  \left[\frac{{\bf r}_{k+2}-{\bf r}_{k+1}}{2Ddt}-\frac{dt}{D\gamma^{2}}\left(\nabla V({\bf r}_{k+1})+\nabla V({\bf r})\right)\right]\label{eq:2}
e^{\left[-\left(\frac{\left({\bf r}_{k+2}-{\bf r}_{k+1}\right)^{2}}{4Ddt}+\frac{\left({\bf r}_{k+1}-{\bf r}\right)^{2}}{4Ddt}+\frac{dt}{D\gamma^{2}}\left(V({\bf r}_{k+1})+V({\bf r})\right)\right)\right]} 
\end{eqnarray}
\end{widetext}
By repeating this procedure, we obtain

\begin{widetext}
\begin{eqnarray}
\nabla Q({\bf r},t) & = & \int_{{\bf r}_{k}={\bf r}}^{{\bf r}_{N}={\bf r}_{f}}d{\bf r}_{k+1}\ldots d{\bf r}_{N-1} 
  \left[\frac{{\bf r}_{f}-{\bf r}_{N-1}}{2Ddt}-\frac{dt}{D\gamma^{2}}\sum_{l=k}^{N-1}\nabla V({\bf r}_{l})\right]
 e^{\left[-\sum_{l=k}^{N-1}\left(\frac{\left({\bf r}_{l+1}-{\bf r}_{l}\right)^{2}}{4Ddt}+\frac{dt}{D\gamma^{2}}V({\bf r}_{l})\right)\right]}\label{eq:N}
\end{eqnarray}
\end{widetext}

By summing these $(N-k)$ equations (\ref{eq:1}), (\ref{eq:2}),...,(\ref{eq:N})
and dividing by $(N-k)$ we obtain

\begin{widetext}
\begin{eqnarray}
\nabla Q({\bf r},t) & = & \frac{{\bf r}_{f}-{\bf r}}{2D(t_{f}-t)}Q({\bf r},t)
 - \frac{1}{D\gamma^{2}}\int_{{\bf r}_{k}={\bf r}}^{{\bf r}_{N}={\bf r}_{f}}d{\bf r}_{k+1}\ldots d{\bf r}_{N-1} e^{\left[-\sum_{l=k+1}^{N-1}\left(\frac{\left({\bf r}_{l+1}-{\bf r}_{l}\right)^{2}}{4Ddt}+\frac{dt}{D\gamma^{2}}V({\bf r}_{l})\right)\right]}\nonumber \\
 & \times & \frac{1}{N-k}\Bigg((N-k)dt\nabla V({\bf r})+(N-k-1)dt\nabla V({\bf r}_{k+1})+\cdots+dt\nabla V({\bf r}_{N-1})\Bigg)\nonumber \\
\label{eq:sum}
\end{eqnarray}
\end{widetext}
Taking the continuous limit of eq.(\ref{eq:sum}) yields

\begin{eqnarray*}
2D\nabla\ln Q({\bf r},t) & = & \frac{{\bf r}_{f}-{\bf r}}{t_{f}-t}-\frac{2}{\gamma^{2}}\int_{t}^{t_{f}}d\tau\left(\frac{t_{f}-\tau}{t_{f}-t}\right)\langle\nabla V({\bf r}(\tau))\rangle
\end{eqnarray*}
where the average $\langle \ldots\rangle$ is done over all Langevin paths starting at $({\bf r},t)$
and ending at $({\bf r}_{f},t_{f})$

\begin{eqnarray}
\langle\nabla V({\bf r}(\tau))\rangle=&&\frac{1}{Q({\bf r},t)}\int_{{\bf r}(t)={\bf r}}^{{\bf r}(t_{f})={\bf r}_{f}}\mathcal{D}{\bf r}(\tau) \nonumber \\
&&e^{-\int_{t}^{t_{f}}d\tau\left(\frac{\dot{{\bf r}}^{2}}{4D}+\frac{1}{D\gamma^{2}}V({\bf r(\tau))}\right)}\nabla V({\bf r}(\tau))\label{eq:fundament}
\end{eqnarray}
The Langevin bridge equation thus becomes

\begin{equation}
\frac{d{\bf r}}{dt}=\frac{{\bf r}_{f}-{\bf r}}{t_{f}-t}-\frac{2}{\gamma^{2}}\int_{t}^{t_{f}}d\tau\left(\frac{t_{f}-\tau}{t_{f}-t}\right)\langle\nabla V({\bf r}(\tau))\rangle+\bm{\eta}(t)\label{eq:exbridge}
\end{equation}
which is eq.(\ref{eq:star}) of the article.

\nocite{*}
\bibliography{onebridge}

\end{document}